\documentclass[aps,prx,a4paper,twocolumn]{revtex4-1}

\usepackage[utf8]{inputenc}
\usepackage{amsmath}
\usepackage{graphicx}

\begin{document}

\title{Cavitation Density of Superfluid Helium-4 around 1 K}
\author{An Qu, A.Trimeche, J.Dupont-roc, J.Grucker, Ph.Jacquier}
\affiliation{Laboratoire Kastler Brossel, ENS-PSL Research University, CNRS, UPMC-Sorbonne-Universités, Collège de France}

\begin{abstract}
Using an optical interferomertric method, the homogeneous cavitation density of superfluid helium at $T=0.96\,$K is measured and found to be $\rho_{cav}=0.1338\pm0.0002\rm\,g.cm^{-3}$. A well established equation of state for liquid helium at negative pressures converts this to the cavitation pressure $P_{cav}=-5.1\pm0.1\,$bar. This cavitation pressure is consistent with a model taking into account the presence of quantized vortices, but disagrees with previously published experimental values of $P_{cav}$.
\end{abstract}

\maketitle

\section{Introduction}
Because helium is a model material at low temperature, its phase transitions have been studied in much detail. In particular, the stability limits of the liquid phase with respect to the solid phase at high pressure and to the gas phase at negative pressure have been studied experimentally and theoretically (see the review articles of S.Balibar \textit{et al.}\cite{balibar2002,balibar2006}). To avoid heterogeneous nucleation of the new phase, experimental over- and under-pressures are produced in bulk liquid away from any surface using high amplitude focused sound waves. While the appearance of the new phase is easily detected optically, the measurement in situ of the local pressure at which nucleation occurs is a challenge. Estimations have been drawn either from the oscillation amplitude of the sound emitter combined with an estimation of the gain due to the focusing \cite{nissen1989,maris1991,caupin2001}, or from extrapolating the liquid static pressure down to where a vanishingly small sound wave would 
produce cavitation \cite{caupin2001}. However nonlinear effects in the sound wave propagation introduce uncertainties in both methods. \par
In 2010, our group introduced a time-resolved quantitative multiphase interferometric imaging technique \cite{souris2010} for measuring the density of a medium inside a sound wave with cylindrical symmetry. In 2012,  this technique was successfully applied for density measurements in solid helium at $1\,$K in pressure swings below the equilibrium melting pressure \cite{souris2011}.  We intended to implement the same technique to study metastable liquid at pressure above this melting pressure. In order to check the method in liquid helium, we decided to measure first the liquid density at which cavitation occurs around $1\,$K aiming at verifying the results of reference \cite{caupin2001}. In this article, we present our result for the cavitation density. Previous experimental results were given as cavitation pressures. So an equation of state for liquid helium at negative pressures have to be used for comparison.  Fortunately, various theoretical approaches \cite{boronat1994,dalfovo1995,bauer2000,maris2002} 
have produced quite similar equations of state. To our surprise, our density result converted to pressure does not quite agree with previous estimations. Thus we also reproduced the pressure extrapolation of F.Caupin and S.Balibar \cite{caupin2001}, and found a reasonable agreement with their data. After discussion of various sources of uncertainties in our measurements, we compare our results with various theoretical estimates of the cavitation pressure.

\section{Experimental set-up}

The technique to measure the local density variations in focused acoustic waves has been described in previous articles \cite{souris2010,souris2011}. Here is a brief reminder.\par
\begin{figure}[!h]
  \centering
    \includegraphics[width=0.5\textwidth]{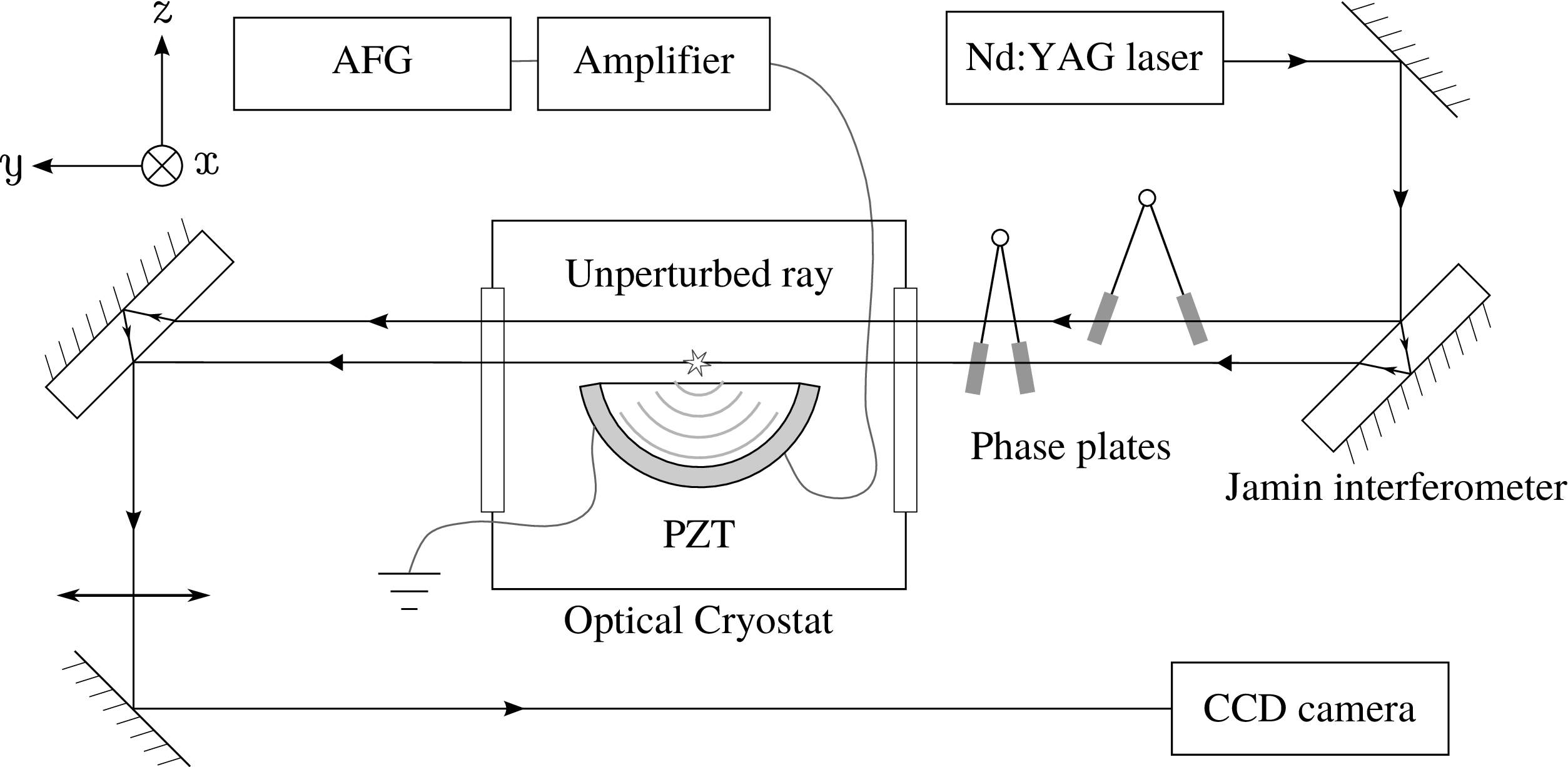}
\caption{Experimental set-up. At the center of the transducer, the small star is the acoustic focus where cavitation occurs. CCD camera monitors the focus plane through a lens. The camera image is composed of pixels distributed in lines and columns. We optimized the camera orientation so that lines are parallel to the $x$-axis.
\label{fig:apparatus}}
\end{figure}

The experimental cell containing liquid helium is cooled in the cryostat with four optical ports. The working temperature can be regulated from 0.9 K to 2.1 K. In the experiment described in this paper, the temperature is fixed at $T=0.96\,$K. The cell is connected to a buffer volume at room temperature, so that the static pressure $P_{st}$ is easily monitored with the help of a Keller X35 pressure sensor with an accuracy of $\pm15\,$mbar. 
A hemispherical piezoelectric transducer (PZT) excites and focuses ultrasound waves in helium at the frequency $1.15\,$MHz of its first thickness vibration mode. The transducer inner diameter is $12\,$mm, and the thickness is $2\,$mm. In order to observe directly the acoustic focus, a small part, $0.9\,$mm in height, has been removed around the transducer rim. One side of the transducer is grounded and the other side is connected to the output of a RF amplifier driven by an arbitrary function generator (AFG). A detailed scheme is shown in Fig.\ref{fig:apparatus}. \par
The optical part of the set-up is a Jamin interferometric imaging system with a pulsed Nd:YAG laser ($\lambda_o=532~$nm) as a light source and a CCD camera for detection. The focal region is imaged onto the CCD camera with a magnification factor of 2 using an appropriate lens. The spatial resolution of the entire system is $20~\mu$m \cite{souris2010}, about $1/10$ of sound wavelength $\lambda_s=0.24\,$mm. One arm of the interferometer passes through the acoustic focal region while the other crosses the cell in an unperturbed zone. The acoustic wave introduces a density variation that modulates the refractive index of helium (mainly in the acoustic focus) and hence gives rise to an optical phase shift between the two paths. A pair of phase plates is placed between the laser and the cell. One of these phase plates is mechanically controlled by a computer in order to add a known phase shift to the unperturbed ray. This added phase shift enables us to extract the phase shift due to the variation of the 
refractive index as we will see in the next section.\par

\section{Local density measurement}

Using the AFG, we can choose the time $t$ with respect to the sound pulse triggering at which an image of the interference field is taken by the camera. This time is adjusted by steps of $0.05~\mu$s, about 6\% of the sound period. By repeating the measurements and recording images at successive delays on a relatively long period, we are able to reconstruct a temporal evolution of the interference field in the focal region.\par

The observed intensity of a given pixel at time $t$ is expected to be:
\begin{equation}
 I(t)=I_0(t)[1+C(t)\cos(\delta \phi(t)+\beta)]
 \label{intensity}
\end{equation}
$\delta\phi(t)$ is the optical phase to be measured, and $\beta$ is the controllable phase shift due to the phase plates. $I_0(t)$ and $C(t)$ are respectively the mean intensity and the fringe contrast. A series of measurements with different $\beta$ at the same time enables us to extract the phases $\delta\phi(t)$ through a fit\cite{souris2010} of eq.(\ref{intensity}).\par
Let $y$ be the light propagation axis, $z$ the PZT axis and $x$ the axis orthogonal to $y$ and $z$ (Fig.\ref{fig:apparatus}).  At a given delay $t$, the phase shift $\delta \phi(x,z)$ is related to the refractive index map by a simple integration along the $y$-axis. In our case, the sound field is rotationally invariant around the hemisphere axis $z$, so that the refractive index variation $\delta n$ is only a function of $z$ and $r=\sqrt{x^2+y^2}$. Given the fact that $\delta n$ is $0$ outside sound field, we can write $\delta\phi$ as the Abel transform of $\delta n$:
\begin{equation}
\delta \phi(x,z)=\frac{2\pi}{\lambda}\int_{-\infty}^{+\infty}dy\,\delta n(\sqrt{x^2+y^2},z)
\end{equation}
Conversely, radial refractive index profiles can be retrieved from phase shift maps via an inverse Abel transform. Then, using Clausius-Mossotti relation in the limit $n_0\sim1$, the density variation $\delta\rho$  of the medium can be deduced easily:
\begin{equation}
 \frac{\delta n}{n_0-1}=\frac{\delta\rho}{\rho_0}
 \label{index}
\end{equation}
$n_0$ and $\rho_0$ being the unperturbed refractive index and density.

\section{Determining the cavitation voltage}

Our measurement of local density by interferometry requires a completely reproducible phenomenon. If bubbles appear randomly, it is impossible to measure the optical phase shift $\delta \phi(t)$ for the pixels involved. In other words, this method only allows us to measure the local density just below the cavitation density where no cavitation process occurs (or the cavitation probability is very low). Then the measured local density should be very close to the real cavitation density and a linear extrapolation to the cavitation voltage would introduce only a small correction. 

\begin{figure}[!h]
  \centering
    \includegraphics[width=0.48\textwidth]{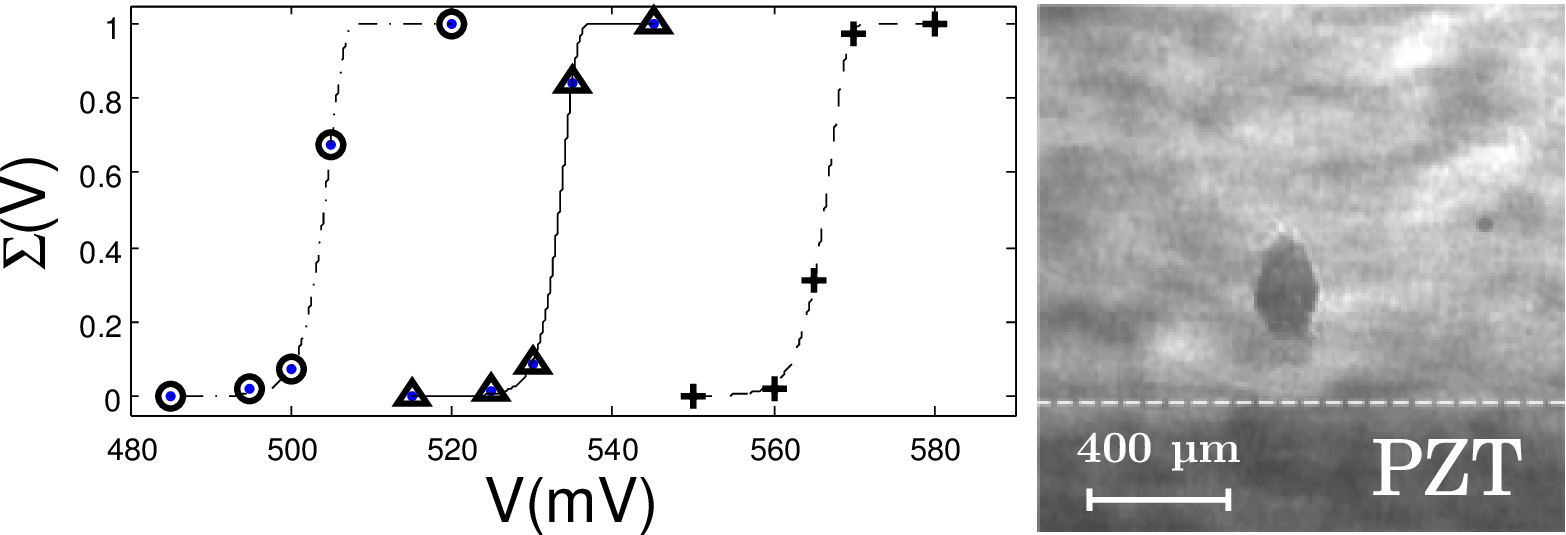}
\caption{Left: Cavitation probability at $0.96\,$K for three different pressures $P_{st}$: circles $0.15\,$bar, triangles $0.65\,$bar, crosses $1.26\,$bar. The corresponding lines are fits according to eq.(\ref{eq:scurve}). Right: Image of a bubble recorded by the camera $10\rm\mu s$ after its creation.
\label{fig:The-probability-S}}
\end{figure}

Thus, before performing any density measurement, we have to precisely determine the cavitation voltage. Other groups have observed that bubble life time in superfluid helium depends on the static pressure $P_{st}$ and is of the order of some tens of microseconds \cite{nissen1989,balibar1995}. Then, $\sim 10\,\mu$s after the minimum pressure wave front passed the acoustic focus, bubbles have expanded to their maximum size and are easily observed on the CCD camera (see Fig.\ref{fig:The-probability-S}).

The cavitation process has a statistical behaviour because of the thermal fluctuations. According to F.Caupin \textit{et al.}\cite{caupin2001}, this probability is described by the ``asymmetric S-curve formula'':
\begin{equation}
 \Sigma(V)=1-\exp[-\ln2\exp(\xi(V/V_c-1))]
 \label{eq:scurve}
\end{equation}
where $V$ is the excitation voltage, $V_c$ the cavitation voltage and $\xi$ a dimensionless parameter.
In order to determine $V_c$, we proceed as follows. For a given static pressure, the bubble probability is determined for 5 different excitation voltages. Each voltage point corresponds to 1000~trials (1000~sound pulses) and the probability is then simply given by the number of positive events (creation of bubble) divided by the number of trials. The relative standard deviation on the probability is $1/\sqrt{1000} \simeq3\%$. To avoid heating, 10 bursts of 100 sound pulses at 10~Hz repetition rate were shot, waiting 100~s between each burst.
To precisely control the driving voltage of the PZT, we fixed the RF amplifier gain factor at about $390$, and adjust only the AFG voltage amplitude with a relative accuracy of 10$^{-4}$. Hence we use the AFG voltage~$V$ as a scale to determine the cavitation voltage instead of the PZT driving voltage.

The cavitation voltage is the value corresponding to a bubble probability of 1/2 according to eq.\ref{eq:scurve}. As it can be seen in Fig.\ref{fig:The-probability-S}, the relative width of the curves is about $1$\% of $V_c$. These curves are indeed very sharp and for the AFG voltage $V_{max}$ of about $2$\% below $V_c$, the probability $\Sigma(V_{max})$ is about 10$^{-3}$.\par

\section{Cavitation density}

\begin{figure}[!h]
 \centering
  \includegraphics[width=0.45\textwidth]{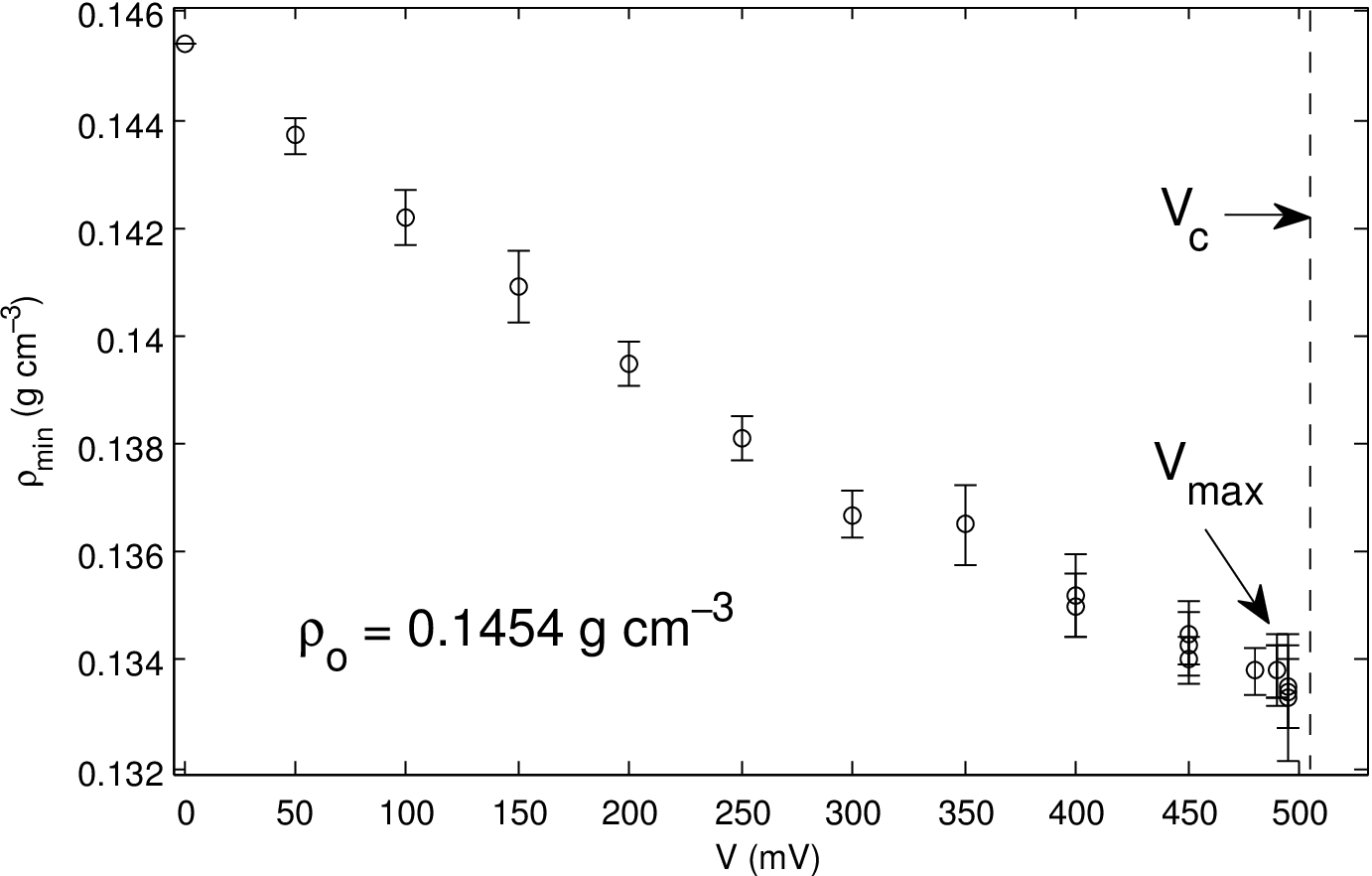}
\caption{Experimental measurements of minimum densities $\rho_{min}$ for different driving voltages $V$ at $T=0.96\,$K and $P_{st}=0.15\,$bar.
The dashed line represents the cavitation voltage.}
\label{fig:variation de densite 152mbar}
\end{figure}

In this section, we present our results for the
cavitation density of superfluid helium at $T=0.96$
K. The cavitation density was reached from three different static pressures
: $0.15$ bar, $0.65$ bar and $1.26$ bar. For each static pressure, 
the cavitation voltage is determined following the approach depicted
previously (see Fig.\ref{fig:The-probability-S}). Then the minimum density in time and
space $\rho_{min}$ was measured for several voltages below the cavitation
threshold. An example is shown in Fig.\ref{fig:variation de densite 152mbar}.
The density variation is not a linear function of the voltage.
This is due to nonlinear effects in the acoustic wave and possible
to the appearance of shock waves \cite{appert2003}.
Assuming a local linear dependence of $\rho_{min}$ to $V$, and taking into account the error bars, one can safely
consider $\rho_{cav}$ is $\rho_{min}(V_{max})$.
In order to improve the accuracy in determining $\rho_{cav}$, we have made a lot of measurements
of the local density just below the cavitation voltage.

The measurement uncertainties can be divided in
two parts : the statistical ones which come mainly from the extraction
process of the optical phase shift induced by the acoustic wave, and
the systematic errors arising from an imperfect cylindrical symmetry
of the pressure wave.

As we mentioned earlier, for each time and for each
pixel, the phase shift value $\delta\phi\left(t\right)$ is obtained
by applying a fit on $\beta$-dependent intensities using eq.(\ref{intensity}) \cite{souris2010}. We use a computer
program to extract these phases with $95\%$ confidence bounds. This
gives the phase shift uncertainties mainly due to shotnoise, camera reading noise and laser power fluctuations. Once the phase shift map is determined, an inverse Abel transform \cite{souris2010} is
applied to recover the refractive index local variation induced by
the acoustic wave. Then, the density variations are deduced from the
optical index variations using eq.(\ref{index}). The inverse Abel transform is a linear transformation. For a given line of an image, the calculation
of the optical index at a given pixel $i$ depends linearly on the
phase shift values for all pixels on the same line. The local
density variation at this pixel is thus in the form :
\begin{equation}
\delta\rho_{i}=\sum_{j=i}^{\infty}\alpha_{ij}\delta\phi_{j}
\label{local_dens}
\end{equation}
where $j$ is the pixel index and $\alpha_{ij}$ is a weight. The errors $\triangle\delta\rho_{i}$ on $\delta\rho_{i}$ can
be computed from the error $\triangle\delta\phi_{j}$ on $\delta\phi_{j}$
and the weights $\alpha_{ij}$ which could be in principle extracted from
the Abel inversion program. Instead we used a simpler empiric method,
assuming that the phase uncertainty is about the same for each pixel,
and is not correlated from one pixel to an other. In that case, the
uncertainty of the density variation at the pixel $i$ is :
\begin{equation}
\left(\triangle\delta\rho_{i}\right)^{2}=\sum_{j=i}^{\infty}\alpha_{ij}^{2}\left(\triangle\delta\phi_{j}\right)^{2}=\left(\triangle\delta\phi\right)^{2}\sum_{j=i}^{\infty}\alpha_{ij}^{2}
\label{uncert_local_dens}
\end{equation}
Then, we performed $N=1000$ density calculations, for the same treated
line while adding a gaussian noise to the phase shifts for every calculation. The
standard deviation of the added noise is chosen to be the same as the phase shift
uncertainty $\triangle\delta\phi$. Once we have these $N$ treatments,
the statistical uncertainty $\left(\triangle\delta\rho_{i}\right)_{N}$
of the radial density variation is calculated for each pixel of the
line. By construction, this uncertainty is equal to $\sqrt{2}$ times
the original unknown statistical uncertainty $\triangle\delta\rho_{i}$
of the density variation, because :
\begin{equation}
\left(\triangle\delta\rho_{i}\right)_{N}^{2}=\sum_{j=i}^{\infty}\alpha_{ij}^{2}\left[\left(\triangle\delta\phi\right)^{2}+\left(\triangle\delta\phi\right)^{2}\right]=2\left(\triangle\delta\rho_{i}\right)^{2}
\label{uncert_stat_dens}
\end{equation}
Appling this method, we found that the statistical uncertainty around the cavitation density is on the order
of $0.0002\,\textrm{g.c\ensuremath{m^{-3}}}$, while the maximum value of $\delta\rho$ is of order $0.0125~\rm g.cm^{-3}$.

The inverse Abel transform assumes that the symmetry
axis is exactly known. Actually, it is unknown and has to be determined
experimentally by searching a symmetry axis in the phase maps. But
the phase noises as well as any possible asymmetry of the acoustic wave locally perturb the left-right symmetry of the phase shift maps.
This perturbation will add an uncertainty in the calculation
of density variations. The difference between the Abel inversion applied
to the left and to the right of this axis gives an order of magnitude
of this uncertainty.

\begin{figure}[!h]
 \centering
  \includegraphics[width=0.45\textwidth]{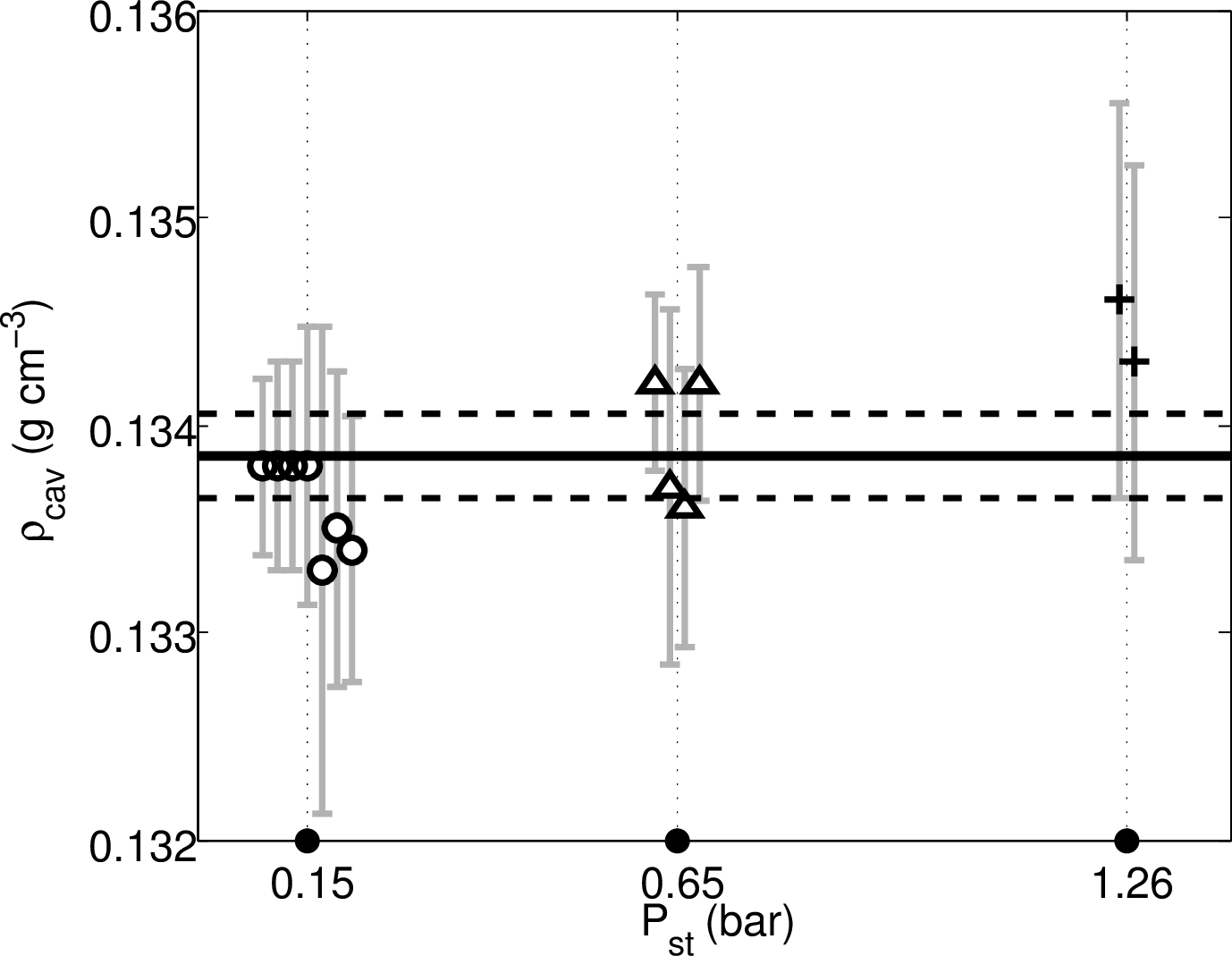}
\caption{Cavitation density as a function of static pressure at $T=0.96\,$K. For more clarity, the different measurements are shifted from their actual $P_{st}$ values (black circles). The horizontal continuous line represents the average cavitation density and the dashed lines its uncertanty. \label{fig:Cavitation-density}}
\end{figure}

The symmetry axis for a given phase map is found
by fitting a straight line through all symmetry centers when the amplitude of the sound pulse at focus is maximum. Then the mean and the
standard deviation for the position of these axis are computed.The uncertainty
on the symmetry axis is about $3\,\textrm{\ensuremath{\mu}m}$,
giving a contribution to the density variation uncertainty on the
order of $0.0003\,\textrm{g.c\ensuremath{m^{-3}}}$. To this systematic
uncertainty we add another incertitude due to the difference between
the left and right parts of the Abel inversion. This gap varies from
one image to another and it is on the order of $0.0005\,\textrm{g.c\ensuremath{m^{-3}}}$. 

We have measured many minimum densities in the vicinity
of the cavitation voltage at several static pressures (0.15 bar, 0.65
bar and 1.26 bar) and at the same temperature $0.96\,$K as shown in Fig.\ref{fig:Cavitation-density}.
In this figure, the error bars represent the quadratic sum of statistical
and systematic uncertainties for each minimum density measurement. Within the error bars we find as expected that $\rho_{cav}$ is independent of $P_{st}$.
Thus, we computed the mean of these measurements and their mean squared
error to determine, respectively, the cavitation density of helium
and its uncertainty. Our final result is that the cavitation density of superfluid
helium-4 at 0.96 K $\rho_{cav}=0.1338\pm0.0002\rm~g.cm^{-3}$.

Note that the uncertainty given here $\left(2\cdot10^{-4}\,\textrm{g.c\ensuremath{m^{-3}}}\right)$
compared to the density variation $\delta\rho$ gives a relative uncertainty about $2\%$.
Concerning the reliability of this measurement, it may be interesting to recall that a comparison with a hydrophone \cite{souris2010} was made in water in 2010. It was found that the deviation between the two methods
is less than 5\%.

\section{Discussion}

Previous results \cite{caupin2001,maris1991} about cavitation in liquid helium were given as cavitation pressures instead of cavitation densities. The equation of state (EOS) of liquid helium in its metastable state (density and pressure below the boiling curve values) is needed to convert the $\rho_{cav}$ to a corresponding $P_{cav}$. Although such an equation of state has never been measured experimentally, some have been proposed. H. Maris has pointed out that, in the stable phase at $T=0.1~K$, the sound velocity pressure dependence could be fit very well by the law $c^3=b(P-P_c)$ with $c$ the sound velocity, $P$ the pressure, $P_c$ the spinodal pressure and $b$ a constant \cite{maris1995}. He proposed that this relationship holds in the metastable state (negative pressure). Bauer \textit{et al.} have
performed Path-integral Monte Carlo simulations of liquid helium in the metastable state at finite temperature and found 
the same dependence of sound velocity on pressure\cite{bauer2000}. Dalfovo \textit{et al.} have calculated the EOS of
metastable liquid helium at $T=0~K$ using density-functional approach\cite{dalfovo1995}, and Boronat \textit{et al.}
using a quadratic diffusion Monte Carlo method to achieve a similar EOS\cite{boronat1994}. 
The EOSs at $0~K$ agree within a few percent. Moreover, using the density-functional theory of Dalfovo \textit{et al.},  Maris and Edwards have shown that in the temperature range $0<T<1~K$, the EOS is nearly independent of temperature\cite{maris2002}.

So in order to compare our cavitation density result to cavitation pressure results of other experiments, we use the well established EOS of metastable liquid helium at $T=0~K$ and assume it holds for $T=0.96~K$. By doing so, our cavitation pressure is $P_{cav}(0.96K)=-5.1\pm0.1\,$bar. 

At temperatures $\sim$1~K, in addition to the present experiment, there are to our knowledge only two experiments which studied the cavitation of liquid helium. Both also used focused acoustic wave. Xiong \textit{et al.}\cite{maris1991} found the cavitation pressure at $1~K$ is $\sim-3\,$bar. The incertitude mentioned in this paper is about $\pm10$\% and comes mostly from the difficulty of estimating the pressure at acoustic focus knowing the displacement of the emitter. Non-linear effects were not taken into account in their calculation. So this incertitude is likely to be underestimated. Caupin \textit{et al.}\cite{caupin2001} studied the dependence of cavitation voltage to the static pressure. They claim that this method enables them to set an upper limit for the actual cavitation pressure. Modelling a linear response of their emitter to voltage, they also give a lower limit for the cavitation pressure. Their result is $-9.8<P_{cav}(0.9K)<-7.7\,$bar. According to the data points published in \cite{
caupin2003}, the result at 
$T\sim$1~K is almost the same. One can see that there are large discrepancies among these experiments.

We have tried to reproduce the experiment of F.Caupin \textit{et al.} using their extrapolation method on $P_{st}$ \cite{caupin2001} (see appendix). The upper limit of $P_{cav}$ we found is about $-8\,$bar which agrees pretty well to 
the one of F.Caupin \textit{et al.}. But the disagreement with our density measurement converted to pressure remains.

Jezek \textit{et al.}\cite{jezek1993} have calculated the cavitation pressure of liquid helium as a function of temperature, by using a density functional method and assuming the absence of defects (especially vortices). In order to compute the cavitation pressure, the volume $\upsilon$ and the time $\tau$ in which nucleation is likely to occur are needed. We take $\upsilon=(\lambda_{s}/2)^3$ and $\tau=0.1\:\mu$s is the $1/10$ of the sound period. This gives $\upsilon\tau\sim\rm10^{-13}~cm^3s$. Using this $\upsilon\tau$ value, Jezek \textit{et al.} calculated $P_{cav}^{Jezek}(0.96K)\sim-6.9\,$bar. This value is just between our result ($-5.1$ bar) and the central value ($-8.8$ bar) of reference\cite{caupin2001}.

Finally, we would like to point out that Maris has developed a model of cavitation in the presence of quantized vortices in liquid helium \cite{maris1994}. For a vortex density ranging from $10^4$ to $10^{12}\,\rm cm^{-2}$, he founds that $-5.8<P_{cav}^{vortices}(0.96K)<-5.1\,$bar. Although Maris can not estimate the error bar on this simulation, we note that our result does lie in this range. Besides, Pettersen \textit{et al.} \cite{pettersen1994} have proposed that the vortex density in the high amplitude sound wave should be of the order of $10^8\sim 10^{10} \,\rm cm^{-2}$. The presence of vortices might be a possible way to conciliate our experimental result with simulations. However, this would imply that, in the presence of vortices, the $P_{st}$ extrapolation method of reference \cite{caupin2001} does not give an upper limit of $P_{cav}$.
\section{Conclusion}
Using an interferometric set up, we have measured the cavitation density of liquid helium-4 at $T=0.96\,$K and the result is $\rho_{cav}=0.1338\pm0.0002\,$g/cm$^3$. Trying to compare this result with existing calculations on the cavitation pressure, we found that a model taking into account the presence of vortices in the liquid can rather satisfactory explain our result. We plan to investigate the influence of vortices on cavitation in liquid helium in two ways. First by studying the dependence of $\rho_{cav}$ 
on temperature, a signature while crossing the lambda temperature should be seen. Second, we will probe the dependence on the density (in the metastable state) of the sound velocity and of the sound attenuation. This last part will be done using stimulated Brillouin scattering.  
\begin{acknowledgments}
 We thank the Laboratoire Kastler Brossel mechanics workshop led by J.M. Isac for their support and especially O.S. Souramasing in making the PZT holder. Special thanks to S. Balibar and F. Caupin for lending us the RF amplifier and many helpful advices. 
\end{acknowledgments}

\appendix*
\section{Cavitation pressure by extrapolating the static pressure}
In 2001, F.Caupin \textit{et al.} implemented an extrapolation method \cite{caupin2001}
to investigate the behavior of helium in negative pressure. They imagined
an environment with a ``stable'' negative static pressure. In this
situation, the required driving voltage for achieving cavitation would
be less than the one in null static pressure. Then the very negative
static pressure corresponding to zero cavitation voltage would be
the cavitation pressure. That can be expressed as : 
\begin{equation}
P_{cav}=P_{focus}=P_{st}+\Delta P\left(\rho_{st}V_{c}\right)
\end{equation}
where $P_{cav}$, $P_{st}$ and $P_{focus}$ are respectively the
cavitation pressure, the static pressure of helium and the pressure
at acoustic focus. $\Delta P\left(\rho_{st}V\right)$ is the variation
of pressure induced by the sound wave and $\rho_{st}$ is the static
density. This equation holds when the driving voltage $V$ reaches
cavitation voltage $V_{c}$. Assuming that the cavitation pressure
is independent of $P_{st}$, we measure the different cavitation
voltages at different static pressures, and then extrapolate linearly
at zero cavitation voltage \footnote{Although we use the AFG generator voltage, it is strictly proportional
to the real driving voltage so that the extrapolation result will
not be affected.}. Numerical simulations \cite{appert2003} have shown that in absence of vortices the true curve is concave
toward negative pressures. The linear extrapolation gives an upper
limit of the true cavitation pressure.

\begin{figure}[!h]
 \centering
  \includegraphics[width=0.5\textwidth]{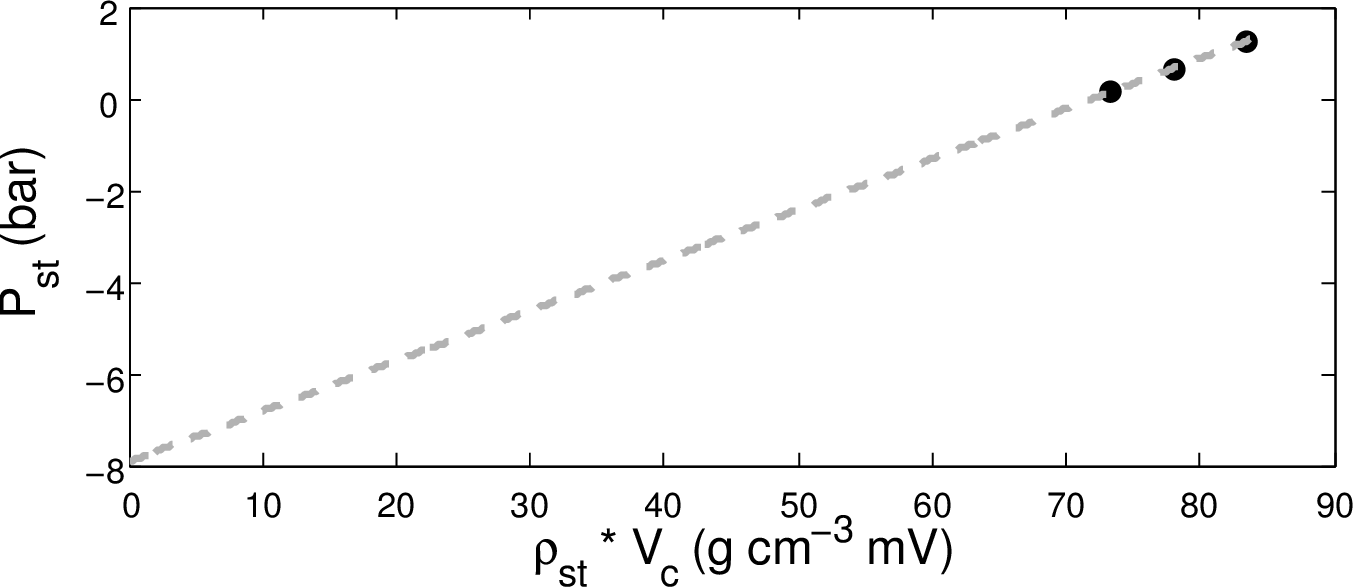}
\caption{Reproduction of reference \cite{caupin2001} experiment. $P_{st}$ as function of $\rho_{st}V_c$.
The 3 data points(black circles) correspond to static pressures of 0.15 bar, 0.65
bar, 1.26 bar. Cavitation voltage are the ones shown in Fig.\ref{fig:The-probability-S}. The dotted line is the linear extrapolation of the date points.
\label{fig:Extrapolation}}
\end{figure}

The Fig.\ref{fig:Extrapolation} shows our extrapolation corresponding to our measurements of ($P_{st}$,$V_c$) values (see Fig.\ref{fig:The-probability-S}).
The upper limit of cavitation pressure obtained in this way is $-7.9 \pm0.3\,$bar.
This is in agreement with F.Caupin \textit{et al.}'s value.

\bibliography{essai.bib}{}
\bibliographystyle{ieeetr}

\end{document}